\documentclass[twocolumn,showpacs,preprintnumbers,amsmath,amssymb,prb,superscriptaddress,aps,10pt]{revtex4-1}

\usepackage{graphicx}

\begin{document}

\title{Signatures in the conductance for phase transitions in excitonic systems}
\author{H. Soller}
\affiliation{Institut f\"ur Theoretische Physik,
Ruprecht-Karls-Universit\"at Heidelberg,\\
 Philosophenweg 19, D-69120 Heidelberg, Germany}
\author{D. Breyel}
\affiliation{Institut f\"ur Theoretische Physik,
Ruprecht-Karls-Universit\"at Heidelberg,\\
 Philosophenweg 19, D-69120 Heidelberg, Germany}

\date{\today}

\begin{abstract}
We analyse two phase transitions in exciton bilayer systems: a topological phase transition to a phase which hosts Majorana fermions and a phase transition to a Wigner crystal. Using generic simple models for the different phases we discuss the conductance properties of the latter when contacted to metallic leads and demonstrate the possibility to observe the different phase transitions by simple conductance measurements.
\end{abstract}

\keywords{excitons, phase transitions, conductance}

\maketitle

\section{Introduction}

Bilayer systems which host excitons represent one of the most interesting class of systems since the bound electron-hole pairs represent a neutral boson while their constituents are charged fermions. If the particles 'feel' their mutual Coulomb attraction but the layers are insulated from each other indirect excitons, consisting of layer-separated electron-hole pairs,  can have extremely long lifetimes.\cite{eisenstein1,snoke,halperin} Due to the high tunability of experimental parameters such systems offer a comparable variability as in cold atom experiments with the additional possibility of direct investigation via electric current.\cite{2012arXiv1206.6626G,2012arXiv1203.3208N,gossard,kuzemsky} Recently increasing attention has been triggered by new experimental results both for the BEC and the BCS phase. However, excitonic bilayer systems offer a much richer variety of phases due to their electronic nature and the electric dipole moment of the electron-hole pair.\cite{PhysRevLett.98.060405}\\
In this paper we want to show that many aspects of this rich variety can be directly detected solely measuring the electric current\cite{2012arXiv1203.3208N} without going over to more elaborate optical readout schemes\cite{gossard,2012arXiv1210.3176A}. Indeed, the different phases of the bilayer system lead to qualitatively different results for the current properties.\\
In this work we will concentrate on two examples: the phase transition to a Wigner crystal-like phase and the topological phase transition to a phase hosting Majorana fermions. While the former has already been discussed to some extent in the literature\cite{springerlink:10.1134/1.1826178} we will show that also the latter exists.\\
The paper is organized as follows: Section \ref{s2} will introduce the basic Hamiltonian formalism we will use in order to describe a non-interacting excitonic system and its transport properties. The topological phase transition in this system is discussed in Section \ref{s3} along with its consequences for the drag conductance. In Section \ref{s4} we will use our description to investigate the photo-electric effect in an exciton bilayer illuminated by a laser field both with dipole-dipole interactions of the excitons and without. We will summarize our results in Section \ref{s5}.

\section{Model} \label{s2}

\begin{figure}[th]
\centering
\includegraphics[width=6cm]{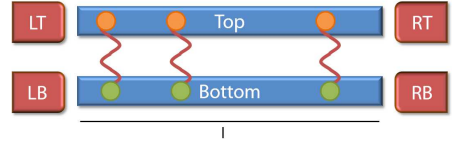}
\caption{Sketch of the experimental setup. The double layer EB is contacted with four metallic leads.}
\label{fig1}
\end{figure}

The basic model system is well established\cite{2013arXiv1301.4182S,PhysRevLett.108.156401} and depicted in Fig. \ref{fig1}: an exciton bilayer (EB) is contacted by four metallic electrodes via tunnel contacts and described by the Hamiltonian
\begin{eqnarray}
H = H_N + H_T + H_{\mathrm{EB}}, \label{htotal}
\end{eqnarray}
where the term $H_N$ describes four metallic drains which are described as fermionic continua written in terms of electron field operators $\alpha_\sigma$ at chemical potentials $\mu_{\alpha\sigma}$ using Fermi distributions $n_{\alpha\sigma}$. $\alpha = L,R$ refers to the contacts on the left/right side of the EB and $\sigma = T,B$ labels the top and bottom layer, respectively. The metallic drains are taken to be in the wide flat band limit so that their density of states (DOS) $\rho_0$ is constant. The additional sum over spin degrees of freedom can be dropped here since either spin is irrelevant as the system is a non-interacting mixture of spins\cite{comte2} or (using a high enough magnetic field) the system is spin-polarized.\cite{PhysRevB.65.235304}
\\
Hopping between the leads and the EB is given by
\begin{eqnarray}
H_T = \sum_{\substack{\sigma = T,B \\ \alpha = L,R}} \gamma_{\alpha \sigma} (\alpha_\sigma^+ \Psi_\sigma + \Psi_\sigma^+ \alpha_\sigma), \label{htunnel}
\end{eqnarray}
using tunneling amplitudes $\gamma_{LT,B}, \; \gamma_{RT,B}$. Finally we describe the EB using a simple one-dimensional model\cite{PhysRevLett.108.156401,PhysRevLett.104.027004}
\begin{eqnarray}
H_{\mathrm{EB}}= \int_{-l/2}^{l/2} \!\!\! dx  \, \,   \Psi^\dagger(x)  \left(\begin{array}{cc} H_{T} & \Delta  \\ \Delta^*  & H_{B} \end{array}\right) \Psi^{}(x) \label{hec} \, ,
\end{eqnarray}
where $l$ is the longitudinal distance between the right and left side of the EB, $\Psi = (\Psi_T, \Psi_B)^T$ is the two-layer spinor and $H_T, H_B$ describe the electron/hole single particle Hamiltonian of the top and bottom layer. Interlayer Coulomb interaction gives rise to an exciton order parameter $\Delta$, which we assume to be constant. Each layer has its own chemical potential $\mu_{EB, \; T/B}$ and we denote the corresponding Fermi distribution by $n_{EB, \sigma}$.\\
We use a Green's function method to calculate the cumulant generating function of charge transfer which gives direct access to the currents and possible higher cumulants through the device in question.\cite{PhysRevB.73.195301} We introduce counting fields for every possible electron-transfer in the system $\mbox{\boldmath$\lambda$} = (\lambda_{LT}, \lambda_{LB}, \lambda_{RT}, \lambda_{RB})$. We use units such that $e = \hbar = k_B = 1$ and $G_0 = 2e^2/h$. The calculation of the CGF as a generalized Keldysh partition function has been carried out before.\cite{2013arXiv1301.4182S,PhysRevLett.108.156401}. Here we only give an approximate result which is correct in the limits of small ($V\ll \Delta$) and high ($V\gg \Delta$) bias but incorporates all relevant features of the CGF we want to discuss here
\begin{eqnarray}
&& \ln \chi_L(\lambda_{LT}, \lambda_{LB}) = 4 \tau \int \frac{d\omega}{2\pi} \nonumber\\
&& \times \left[ \sum_{\sigma = T,B} \ln \left\{1+ T_\sigma(\omega) \left[(e^{i \lambda_{L\sigma}} -1) n_{L\sigma} (1-n_{\mathrm{EB},\sigma}) \right. \right. \right. \nonumber\\
&& \left. \left. + (e^{-i \lambda_{L\sigma}} -1) n_{\mathrm{EB},\sigma} (1-n_{L\sigma})\right]\right\} \theta\left(\frac{|\omega_\sigma| - \Delta}{\Delta}\right) \nonumber\\
&& + \ln \left\{1+ T_A(\omega) \left[(e^{i \lambda_{LT}} e^{-i \lambda_{LB}} -1) n_{LT} (1-n_{LB}) \right. \right. \nonumber\\
&& \left.\left. + (e^{i \lambda_{LB}} e^{-i \lambda_{LT}} -1) n_{LB} (1-n_{LT}) \right]\right\} \nonumber\\
&& \left. \times \theta\left(\frac{\Delta - \max(|\omega_T|, \; |\omega_B|)}{\Delta}\right)\right]. \label{cgf}
\end{eqnarray}
We denote $\omega_{T,B} = \omega - \mu_{\mathrm{EB},\sigma}, \; \sigma=T,B$ and the transmission coefficients are given by
\begin{eqnarray}
T_\sigma(\omega) = \frac{4\tilde{\Gamma}_{L\sigma}}{(1+ \tilde{\Gamma}_{L\sigma})^2} \;\; \mbox{and} \;\; T_A(\omega) = \frac{4\tilde{\Gamma}_A}{(1+ \tilde{\Gamma}_A)^2}.
\end{eqnarray}
The energy-dependent DOS of the EB affects the hybridisations
\begin{eqnarray}
\tilde{\Gamma}_{L\sigma} &=& \frac{\Gamma_{L\sigma} |\omega_\sigma|}{\sqrt{\omega_\sigma^2 - \Delta^2}}, \;\; \mbox{and} \;\; \tilde{\Gamma}_A = \frac{\Gamma_{LT} \Gamma_{LB} \Delta^2}{\sqrt{\Delta^2 - \omega_T^2} \sqrt{\Delta^2 - \omega_B^2}}, \nonumber
\end{eqnarray}
where $\Gamma_{L\sigma} = \pi^2 \rho_{0L\sigma} \rho_{0\mathrm{SC}} \gamma_{L\sigma}^2/2$, with the DOS of the EB given by $\rho_{0\mathrm{EB}}$.\\
The chemical potentials of the two EB layers have to be chosen self-consistently in order to ensure current conservation $\langle I_{L,T/B} \rangle = \langle I_{R,T/B} \rangle$.\cite{lambert} This is most easily done for $\Gamma_{L,T/B} = \Gamma_{R,T/B}$ and $\mu_{L,T/B} = - \mu_{R,T/B} = V_{T/B}/2$ since then $\mu_{EB,T/B} = 0$ fulfills the above self-consistency condition.\cite{PhysRevLett.108.156401}\\
We only give the result for the CGF on the left side since it is related to the CGF on the right side by simply exchanging $\lambda_{L,T/B} = - \lambda_{R,T/B}$.\\
From this result we can calculate the different currents as simple derivatives
\begin{eqnarray}
\langle I_{LT} \rangle = - \frac{i}{\tau} \left.\frac{\partial \chi_L}{\partial \lambda_{LT}}\right|_{\mathbf{\lambda} = 0}, \label{currcgf}
\end{eqnarray}
and likewise for the other currents and higher cumulants.

\section{Majorana fermions} \label{s3}

\begin{figure}[th]
\centering
\includegraphics[width=0.4\textwidth]{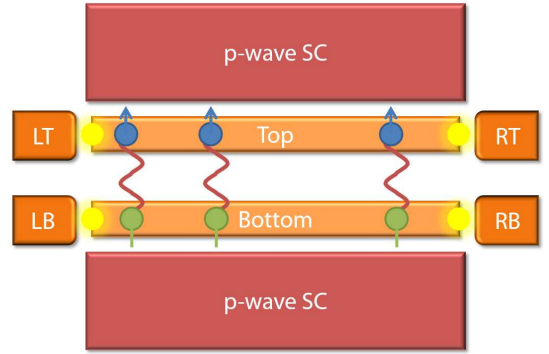}
\vspace*{8pt}
\caption{Setup considered for the generation of Majorana fermions in EBs. A EB is spin-polarized by a magnetic field (indicated by the arrows for the particles in the top and bottom layer). Additionally the EB is proximity coupled to two $p$-wave SCs. The emerging Majorana fermions are indicated as yellow spots in the top and bottom layer.}
\label{fig8}
\end{figure}

One of the most remarkable properties of an EB that can be deduced from Eq. (\ref{cgf}) is perfect Coulomb drag\cite{2012arXiv1206.6626G,su2008} meaning that for $T=0$ and $V_{T/B} < 2 \Delta$
\begin{eqnarray}
\langle I_{LT} \rangle = - \langle I_{LB} \rangle.
\end{eqnarray}
This property has by now been observed experimentally\cite{2012arXiv1203.3208N} and is a typical feature of an EB as described by Eq. (\ref{hec}). Now, we want to observe the change of this behavior after a topological phase transition.
First, we would like to illustrate a possibility of generating Majorana fermions in EBs. We use the setup as depicted in Fig. \ref{fig8}: the bilayer EB is assumed to be spin-polarized. Such a situation may be created e.g. by applying a sufficiently large magnetic field.\cite{PhysRevB.65.235304} Additionally the EB is proximity coupled to two $p$-wave SCs.
\\
We want to show that this system has the same low energy behavior as the Kitaev model.\cite{1063-7869-44-10S-S29} Compared to other proposals involving strongly spin orbit coupled nanowires proximity coupled to superconductors (SCs) in a magnetic field \cite{PhysRevLett.105.177002,PhysRevLett.107.036801} we use the spinless nature of the EB in combination with the $p$-wave SC in order to mimick the behavior of the Kitaev model. The simplest Hamiltonian describing the EB in the presence of the $p$-wave SC reads \cite{Volkov1995261}
\begin{eqnarray}
H_{\mathrm{EB,M}} &=& \int dx \Psi_{\mathrm{EB}}^+(x) {\cal H}_{\mathrm{EB,M}} \Psi_{\mathrm{EB}}(x), \\
\Psi_{\mathrm{EB}}^+ &=& (\Psi_T^+, \Psi_B^+, \Psi_B, \Psi_T),  \; \mbox{where}\nonumber\\
{\cal H}_{\mathrm{EB,M}} &=& \left[\frac{p^2}{2m}\right] \tau_z + \Delta \tau_x + \Delta_p \sigma_x \tau_x,
\end{eqnarray}
where we combined Eq. (\ref{hec}) and the proximity induced $p$-wave coupling $\Delta_p$. The Pauli matrices $\tau$ and $\sigma$ operate in particle-hole and top-bottom space, respectively.\\
One should note that we assume the exciton condensate to be two-dimensional and electrically connected to the superconductors on each side in a sandwich structure. This way we are allowed to treat the proximity induced coupling in a mean-field approach.\cite{Volkov1995261} Furthermore we assume the condensate to be contacted at two points as in typical experiments.\cite{2012arXiv1203.3208N}\\ 
The spectrum can be revealed\cite{PhysRevLett.105.177002,PhysRevLett.107.036801} by squaring ${\cal H}_{\mathrm{EB,M}}$ twice, which yields
\begin{eqnarray}
E_{\mathrm{EB},\pm}^2 = \Delta_p^2 + \Delta^2 + \xi_p^2 \pm (2 \Delta \Delta_p),
\end{eqnarray}
where $\xi_p = p^2/(2m)$. The gap\cite{PhysRevLett.105.177002,PhysRevLett.107.036801} $E_{\mathrm{EB},0}$ near $p=0$ is the key to the emergence of Majorana fermions. We find
\begin{eqnarray}
E_{\mathrm{EB},0} = |\Delta_p - \Delta|.
\end{eqnarray}
Consequently, we observe an exchange field-dominated (or strong interaction induced) gap if $\Delta_p > \Delta$ and hence the EB in this case will host Majorana fermions. The end of the EB can be characterised by a sharp drop of the chemical potential, which closes the gap. Since this transition corresponds to a transition out of the topological phase Majorana fermions will be localized at the EB ends. However, this was to expected since we have basically created a system consisting of two times the Kitaev model. The only difference to the former proposal is that now we will have four Majoranas instead of two, however, in a physically very different system. The situation is slightly different when discussing topological exciton condensates.\cite{2012arXiv1203.6628S}\\
The main features of the Majorana fermions for the system in question are their localisation at the ends of the EB and the presence of four instead of two. Since we have assumed the EB to be contacted at two points we address a single channel in both layers so that we will see only four Majorana fermions forming at the respective contacted points. There are four due to the additional layer degree of freedom of the EB.\\
As discussed above we are allowed to treat the EB in a mean-field approach. If the EB is in the topological phase both layers can therefore be treated as $p$-wave superconductors with an effective $p$-wave gap of $E_{\mathrm{EB},0}$. Majorana fermions alter the characteristics of Andreev reflection at the interface to such superconductors by contributing an additional phase shift.\cite{PhysRevLett.106.057001} The consequence of this additional phase shift is a perfectly transmitting channel at $V_{T/B} = 0$ through the upper and lower layer of the EB.\cite{PhysRevB.82.180516,PhysRevLett.106.057001} However, since conductance through the upper and lower layer is perfect, there can be no transconductance in this case due to unitarity. Therefore the phase transition to a Majorana fermion phase can be easily seen by a vanishing transconductance and a peak in the differential conductance at $V_{T,B} = 0$ through the layer in question. Indeed, these two signatures can be easily observed by simple condutance measurements and a verification of the presence of Majoranas does not require to also measure e.g. the Josephson current.\cite{PhysRevLett.105.177002}

\section{Photo-electric effect} \label{s4}

\begin{figure}[th]
\centering
\includegraphics[width=7cm]{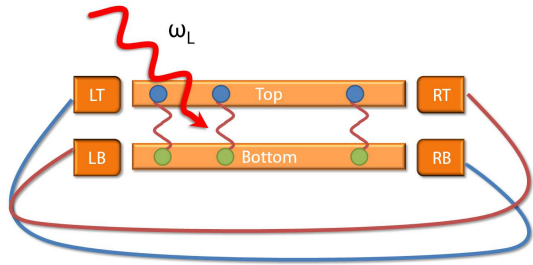}
\caption{Sktech of the setup in order to demonstrate the usage of EBs for solar cells: an exciton condensate is contacted by four metallic electrodes and additionally irradiated by a laser field at frequency $\omega_L$. The different leads are contacted in the way shown in the figure in order to allow for operation as a solar cell.}
\label{fig6}
\end{figure}

\begin{figure}[th]
\centering
\includegraphics[width=0.35\textwidth]{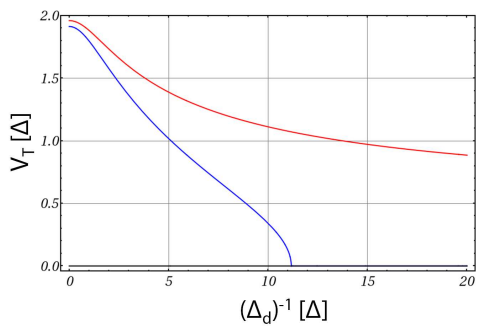}
\vspace*{8pt}
\caption{Self consistent solution of Eq. (\ref{transsolar}) for varying $G_{Ex}$: the induced voltage in the top layer $V_T$ is shown as a function of the detuning $\Delta_d$, both given in units of the EB order parameter $\Delta$=$\Delta_R$. For the EB we choose $\Gamma_{LT} = \Gamma_{RT} = \Gamma_{LB} = \Gamma_{RB} = 0.1$. The red curve refers to $G_{Ex} = 0.005 G_0$, the blue curve is for $G_{Ex} = 0.0045 G_0$ and the black curve is for $G_{Ex} = 0.001 G_0$.}
\label{fig7}
\end{figure}

A possible application of an EB is in the fashion of a solar cell. The most simple setup in order to demonstrate this application, is an EB which is contacted to metallic electrodes and irradiated by a laser field as skteched in Fig. \ref{fig6}. We first study the EB irradiated by the laser field alone and will afterwards introduce the contacts.
We excite the system via a laser field at a frequency $\omega_L$. $\Delta_R$ is the Rabi frequency of the exciton transition and $\Delta_d$ describes the detuning of the laser field from this transition. In a typical experiment one uses short and high intensity laser pulses to excite the system. Under these conditions the Rabi frequency $\Delta_R$ is of the same order as the exciton binding energy.\cite{PhysRevB.48.17811} We therefore set $\Delta_R = \Delta$ in the following to simplify the discussion. $\Delta_d$ is the detuning of the laser field from this transition. Using a laser field allows to generate excitons via exciting an electron from the valence band of the bottom layer. The excitons can now be either excited or not which allows to map the system to a two-level system of $N$ randomly arranged interacting spin-1/2 systems. The Hamiltonian reads\cite{PhysRevLett.98.060405,PhysRevA.86.023405}
\begin{eqnarray}
H_{\mathrm{Ex,Laser}} = - \frac{\Delta_d}{2} \sum_{i=1}^{N} \sigma_z^{(i)} + \frac{\Delta}{2} \sum_{i=1}^N \sigma_x^{(i)}, \label{hfree}
\end{eqnarray}
when we can neglect interactions between the excitons. $\sigma_{x,z}^{(i)}$ denote Pauli matrices in the effective two level system. The model assumes that we use high-intensity laser pulses so that every energetically favorable transition will take place. Using lower laser power will result in a smaller number of excitons.\\
The (generalized) spin-1/2 Ising model in Eq. (\ref{hfree}) is exactly solvable and the expectation value of the number of excitons $N_{\mathrm{ex}}$ is given by
\begin{eqnarray}
N_{\mathrm{Ex}} = \frac{N}{2} \left(1+ \frac{1}{\sqrt{1+ \Delta^2/\Delta_d^2}}\right). \label{nex}
\end{eqnarray}
We may therefore express the exciton density $n_{\mathrm{Ex}}$ by the electron density $n_e$ in the semiconducting material as
\begin{eqnarray}
n_{\mathrm{Ex}} = \frac{1}{2} \left(1+ \frac{1}{\sqrt{1+ \Delta^2/\Delta_d^2}}\right) n_e.
\end{eqnarray}
Therefore we also know the current in the top layer as a function of the applied voltage $V_T$ since
\begin{eqnarray}
\langle I_{\mathrm{Ex}} \rangle = - e n_{\mathrm{Ex}} \mu \frac{W}{l} V_T = G_{\mathrm{Ex}} V_T, \label{currex}
\end{eqnarray}
where $\mu$ is the carrier mobility and $W$ is the width of the bilayer. If, for simplicity, we assume the same carrier mobility in the top and bottom layer the same relation will hold for the bottom layer with an exchanged prefactor and the flow of excitons will create the same electric field, causing the same voltage $V_T = - V_B$ in the two layers. For temperatures $T\ll \Delta$ and $|V_{T,B}| < 2 \Delta$ the current is purely due to excitons travelling though the bilayer. We use the wiring as indicated in Fig. \ref{fig6}. In order to fulfill charge conservation Eq. (\ref{currex}) has to be equal to the current calculated from Eq. (\ref{currcgf}). We therefore have to solve the following equation self-consistently
\begin{eqnarray}
\langle I_{LT} \rangle = \langle I_{\mathrm{Ex}} \rangle. \label{transsolar}
\end{eqnarray}
We can use typical parameters \cite{2012arXiv1203.3208N} $l = B = 1\, \mu$m for the junction, $n_e \approx 10^{11}/$cm$^2$ and $\mu = 100\,$m$^2$V$^{-1}$s$^{-1}$ and typically arrive at rather small conductances $G_{\mathrm{Ex}} \approx 10^{-5} G_0$. However, since the conductance depends on the choice of materials we show the result in Fig. \ref{fig7} for varying $G_{\mathrm{Ex}}$.
\\
Of course, Eq. (\ref{transsolar}) has a trivial solution, meaning $V_T = 0$. However, we observe in Fig. \ref{fig7} that indeed it is possible to have solution where $V_T \neq 0$ which corresponds to a current caused by the excitation of the laser. The latter solution has to be stable since if it were not stable further stable solutions at voltages $V_{T,B} = \pm \infty$ would have to exist which is clearly not the case.\cite{PhysRev.124.41} The voltage depends on the detuning and is generally highest if the detuning is maximal.\\
Therefore, EBs can be operated as solar cells. However, when we assumed Eq. (\ref{hfree}) to describe the EB we neglected interactions between the excitons. If we would additionally include interaction we would have to add a term
\begin{eqnarray}
H_{\mathrm{int}} = \frac{C}{4} \sum_{i=2}^{N} \sum_{j=1}^{i-1} \frac{(1+ \sigma_z^{(i)})(1+ \sigma_z^{(j)})}{|\mathbf{r}_i - \mathbf{r}_j|^3}, \label{hint}
\end{eqnarray}
in which $C$ describes the strength of the mutual dipole-dipole repulsion.\cite{PhysRevA.86.023405,PhysRevLett.98.060405} The inclusion of such a term leads to the formation of a Wigner crystal of the excitons at strong enough interaction $C$. The strength of the interaction needed has been discussed in the literature using different numerical procedures.\cite{PhysRevLett.98.060405,springerlink:10.1134/1.1826178} The formation of such a Wigner crystal at a certain interaction strength $C^*$ leads to crystallization of of the electrons themselves and therefore the number of free charge carriers as calculated in Eq. (\ref{nex}) will be zero. Consequently, we will only encounter the trivial solution of Eq. (\ref{transsolar}) and will observe no photoelectric effect. As a function of the interaction we will therefore observe a vanishing a photoelectric effect when increasing the interaction beyond $C^*$. Consequently, this vanishing photoelectric effect is a clear signature for the phase transition to a Wigner crystal and can again be monitored by simple conductance measurements.\\
We should add that in realistic implementations of the proposed setup tunneling between the two layers cannot be fully suppressed, as assumed here. The role of such a term has been discussed in the literature\cite{su2008,2012arXiv1203.3208N,lozovikyud,2013arXiv1301.4182S} and depends on the specific implementation. In the typical quantum Hall bilayer systems\cite{2012arXiv1203.3208N} the experimental results are in good agreement with a vanishing tunneling term\cite{2013arXiv1301.4182S} so that the above description should be applicable. We also checked that a long but finite exciton lifetime leading to a small imaginary part of $\Delta$ does not alter our results significantly. The situation is different for graphene bilayers where the effect of a tunneling term depends on the nature of the insulating barrier and on the rotation of the graphene layers with respect to each other.

\section{Conclusions} \label{s5}

We have considered two phase transitions in exciton bilayer systems: the topological phase transition to a phase which hosts Majorana fermions and the phase transition to a Wigner crystal. In both cases these phase transitions severly change conductance properties leading either to a vanishing transconductance or a vanishing photo-electric effect. We believe that this scheme can be extended to other phase transitions in excitonic systems which should leave clear signatures in the conductance properties.

\section*{Acknowledgments}

The authors would like to thank A. Komnik, T. L. Schmidt and F. Dolcini for many helpful discussions.

\end{document}